\documentclass[twocolumn,aps,prb,superscriptaddress,address,amsmath,amssymb,showpacs]{revtex4}
\usepackage{amsmath}
\usepackage{amssymb}
\usepackage{graphicx}
\usepackage{esint}
\usepackage[unicode=true,pdfusetitle,
 bookmarks=true,bookmarksnumbered=false,bookmarksopen=false,
 breaklinks=false,pdfborder={0 0 1},backref=false,colorlinks=false]
 {hyperref}
\usepackage{breakurl}

\makeatletter
\@ifundefined{textcolor}{}
{%
 \definecolor{BLACK}{gray}{0}
 \definecolor{WHITE}{gray}{1}
 \definecolor{RED}{rgb}{1,0,0}
 \definecolor{GREEN}{rgb}{0,1,0}
 \definecolor{BLUE}{rgb}{0,0,1}
 \definecolor{CYAN}{cmyk}{1,0,0,0}
 \definecolor{MAGENTA}{cmyk}{0,1,0,0}
 \definecolor{YELLOW}{cmyk}{0,0,1,0}
}

\usepackage{subfigure}
\usepackage{color}

\newcommand{\gtwid}{\mathrel{\raise.3ex\hbox{$>$\kern-.75em\lower1ex\hbox{$\sim$}}}}
\newcommand{\ltwid}{\mathrel{\raise.3ex\hbox{$<$\kern-.75em\lower1ex\hbox{$\sim$}}}}
\makeatother

\begin{document}


\title{Tunneling magnetoresistance devices based on topological insulators:
Ferromagnet/insulator/topological-insulator junctions employing Bi$_{2}$Se$_{3}$}

\author{Matthias G\"otte, Tomi Paananen, G\"unter Reiss, and Thomas Dahm}

\affiliation{Universit\"at Bielefeld, Fakult\"at f\"ur Physik, Postfach 100131, D-33501
Bielefeld, Germany}

\date{\today}
\begin{abstract}
We theoretically investigate tunneling magnetoresistance (TMR) devices,
which are probing the spin-momentum coupled nature of surface states
of the three-dimensional topological insulator Bi$_{2}$Se$_{3}$.
Theoretical calculations are performed based on a realistic tight-binding
model for Bi$_{2}$Se$_{3}$. We study both three dimensional devices,
which exploit the surface states of Bi$_{2}$Se$_{3}$, as well as
two-dimensional devices, which exploit the edge states of thin Bi$_{2}$Se$_{3}$
strips. We demonstrate that the material properties of Bi$_{2}$Se$_{3}$
allow a TMR ratio at room temperature of the order of 1000~\%.
Analytical formulas are derived that allow a quick estimate
of the achievable TMR ratio in these devices.
The devices can be used to measure the spin polarization of
the topological surface states as an alternative to spin-ARPES. 
Unlike TMR devices based on magnetic tunnel junctions the present
devices avoid the use of a second ferromagnetic electrode
whose magnetization needs to be pinned.
\end{abstract}
\maketitle

\section{Introduction}

Topological insulators (TI) are materials which possess an insulating
gap in the bulk but conducting states at the surface. These surface
states are protected by the topological properties of the material
and are robust against time-reversal-symmetric perturbations like
non-magnetic impurities, for example. After first being
predicted theoretically, \cite{Bernevig,Fu} subsequently several
materials have been confirmed to be topological insulators experimentally. 
\cite{Koenig,Hsieh1,Chen,Xia,Hsieh2,Kuroda,Ando} 
Due to spin-orbit coupling the
momentum of the surface states is locked with the spin of the electrons,
which means that electrons with opposite spin propagate
into opposite direction \cite{Ando,Hsieh,Bruene,Pan:PRL106}.
The robustness of the surface states against backscattering promises
long spin diffusion lengths. These features make topological insulators
particularly interesting for applications in spintronics 
\cite{Tanaka,Mondal,Linder,Garate,Yokoyama,BlackSchaffer,Krueckl,Kong,Salehi,Vali,PGGD,Taguchi}.
In addition, spin-dependent tunneling into the surface states opens
a way to investigate their properties such as spin-polarization.

In the present work we theoretically study different arrangements
to realize a tunneling magnetoresistance (TMR) device using a ferromagnet
coupled via a tunnel barrier to a topological insulator. In contrast
to conventional TMR junctions such a device does not need a second
ferromagnetic layer, because the spin locking in the topological insulator
already provides an intrinsic magnetic reference. Additionally, the device
can be used to probe the spin-locked surface states in
the TI. We will show that the material properties of the topological
insulator determine the maximum TMR ratio that can be achieved. Specifically,
we will present calculations for the three-dimensional topological
insulator Bi$_{2}$Se$_{3}$, which has been studied well in the past
both theoretically and experimentally \cite{Zhang:NPhys09,Xia,Liu:PRB10,Li:NPhys10,Shan}.
This material is particularly interesting due to its comparatively
large band gap of about 0.3~eV. As we will see, this allows to achieve
large TMR ratios already at room temperature.

\section{Model}

For our calculations we are using a realistic tight-binding model
for Bi$_{2}$Se$_{3}$ that has been derived from bandstructure calculations
by Liu et al \cite{Liu:PRB10} based on ${\mathbf{k}\cdot\mathbf{p}}$
theory. The Hamiltonian takes into account two orbitals at each lattice
site and reads 
\begin{equation}
H(\mathbf{k})=\epsilon_{0}(\mathbf{k})\mathbb{I}_{4\times4}+\sum_{i=1}^{4}m_{i}\left(\mathbf{k}\right)\Gamma^{i}+\mathcal{R}_{1}\left(\mathbf{k}\right)\Gamma^{5}+\mathcal{R}_{2}\left(\mathbf{k}\right)\Gamma^{3}\label{eq:hamiltonian}
\end{equation}
Here, 
\begin{eqnarray}
\epsilon_{0}(\mathbf{k}) & = & C_{0}+2C_{1}\left(1-\cos k_{z}\right)\nonumber \\
 &  & +\frac{4}{3}C_{2}\left(3-2\cos\frac{1}{2}k_{x}\cos\frac{\sqrt{3}}{2}k_{y}-\cos k_{x}\right)\nonumber \\
m_{1}(\mathbf{k}) & = & A_{0}\frac{2}{\sqrt{3}}\cos\frac{1}{2}k_{x}\sin\frac{\sqrt{3}}{2}k_{y}\nonumber \\
m_{2}(\mathbf{k}) & = & -A_{0}\frac{2}{3}\left(\sin\frac{1}{2}k_{x}\cos\frac{\sqrt{3}}{2}k_{y}+\sin k_{x}\right)\nonumber \\
m_{3}(\mathbf{k}) & = & B_{0}\sin k_{z}\nonumber \\
m_{4}(\mathbf{k}) & = & M_{0}+2M_{1}\left(1-\cos k_{z}\right)\nonumber \\
 &  & +\frac{4}{3}M_{2}\left(3-2\cos\frac{1}{2}k_{x}\cos\frac{\sqrt{3}}{2}k_{y}-\cos k_{x}\right)\nonumber 
\end{eqnarray}
are tight-binding parameters defined on a bilayer hexagonal lattice
following Hao and Lee \cite{Hao}. The terms 
\begin{eqnarray}
\label{termr1}
\mathcal{R}_{1}\left(\mathbf{k}\right) & = & 2R_{1}\left(\cos\sqrt{3}k_{y}-\cos k_{x}\right)\sin k_{x}\\
\label{termr2}
\mathcal{R}_{2}\left(\mathbf{k}\right) & = & \frac{16}{3\sqrt{3}}R_{2}\left(\cos\frac{\sqrt{3}}{2}k_{y}-\cos\frac{3}{2}k_{x}\right)\sin\frac{\sqrt{3}}{2}k_{y}
\end{eqnarray}
are third order terms to lowest order in momentum $\mathbf{k}$ \cite{Liu:PRB10}.
The Dirac $\Gamma$ matrices are represented by $\Gamma^{1,2,3,4,5}=\left(\tau_{1}\otimes\sigma_{1},\tau_{1}\otimes\sigma_{2},\tau_{2}\otimes\mathbb{I}_{2\times2},\tau_{3}\otimes\mathbb{I}_{2\times2},\tau_{1}\otimes\sigma_{3}\right)$,
where $\tau_{i}$ and $\sigma_{i}$ are the Pauli matrices in the
orbital and spin space, respectively. The model parameters are derived
from Liu et al \cite{Liu:PRB10} using the atomic distances $a=4.14\textrm{\AA}$
and $c=\frac{28.64}{15}\textrm{\AA}$ ($a$ and $15 c$ are
the lattice constants given in Refs.~\onlinecite{Okamoto,Lind,Zhang:APL09};
a unit cell consists of 15 layers in c-direction): $A_{0}=0.804\textrm{eV}$,
$B_{0}=1.184\textrm{eV}$, $C_{1}=1.575\textrm{eV}$, $C_{2}=1.774\textrm{eV}$,
$M_{0}=-0.28\textrm{eV}$, $M_{1}=1.882\textrm{eV}$, $M_{2}=2.596\textrm{eV}$,
$R_{1}=0.713\textrm{eV}$, and $R_{2}=-1.597\textrm{eV}$. $C_{0}$
is only a tiny energy shift and is chosen to be $0$, which corresponds
to undoped Bi$_2$Se$_3$. Here and in
the following all energies are given with respect to the Fermi level
of the topological insulator.

\begin{figure}[t]
\centering \includegraphics[width=0.9\columnwidth]{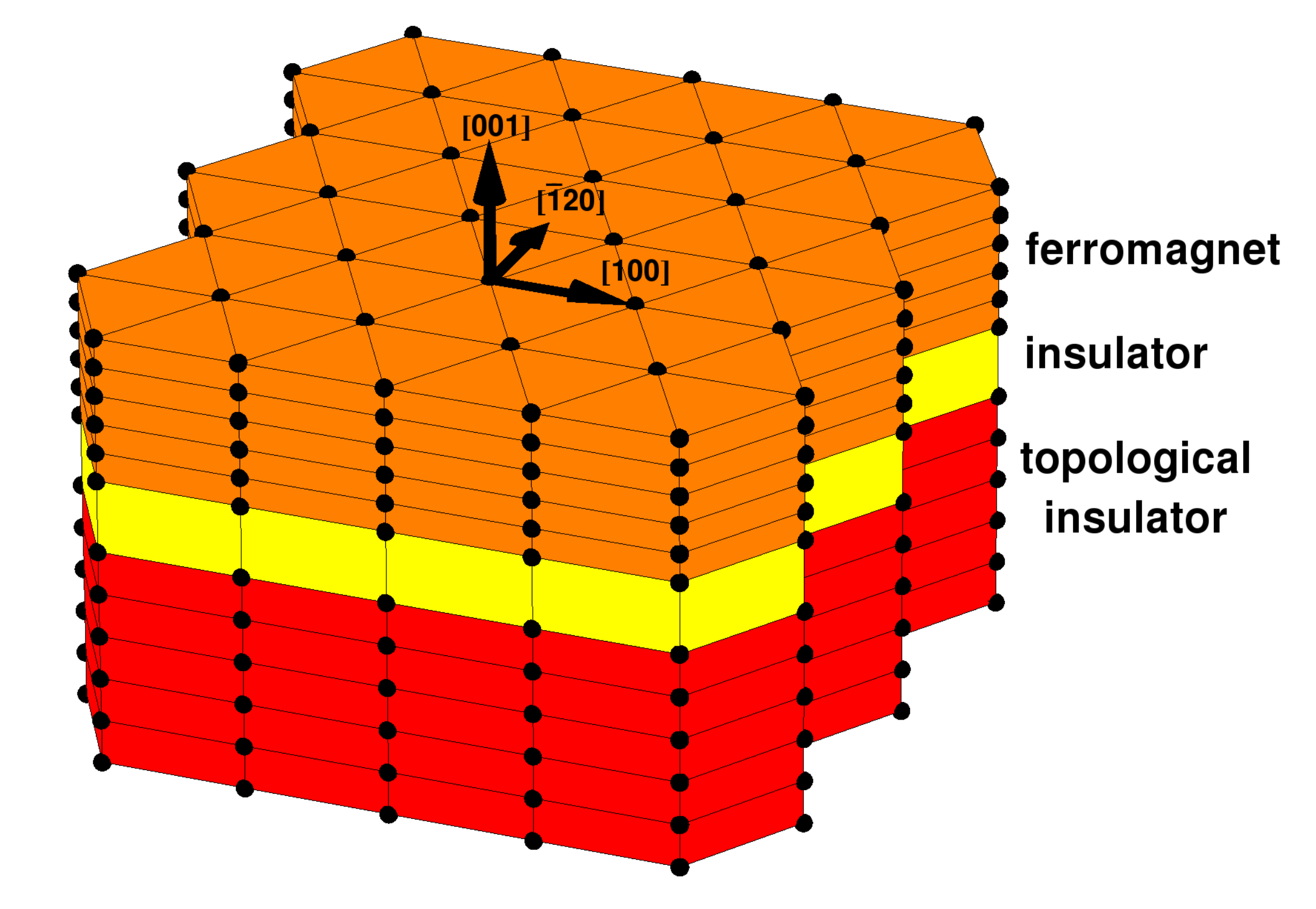} \caption{\label{fig1} Tight-binding model of a ferromagnet-insulator-topological
insulator (FITI) junction with the ferromagnet on the {[}001{]}-surface
of the topological insulator, corresponding to setup A and B in Fig.~\ref{fig2}.}
\end{figure}

In Fig.~\ref{fig1} a tight-binding model of a ferromagnet-insulator-topological
insulator (FITI) junction is shown. The topological insulator is modeled
by the Hamiltonian Eq. \eqref{eq:hamiltonian}, which we Fourier-transform
perpendicular to the junction plane into real space onto its lattice.
For the in-plane directions we assume periodic boundary conditions,
allowing us to keep the in-plane momentum components (e.g. $k_{x}$
and $k_{y}$ for a $z$-surface) as good quantum numbers. The ferromagnet
(FM) is modeled as a metal with two spin-split subbands. For that
purpose we use the same Hamiltonian Eq. \eqref{eq:hamiltonian} with
parameters $C_{F}\equiv C_{1}=C_{2}=0.25\textrm{eV}$ and $A_{0}=B_{0}=M_{0}=M_{1}=M_{2}=R_{1}=R_{2}=0$
corresponding to a 3~eV bandwidth. Additionally we add an exchange
field of strength $W$ of the form 
\begin{equation}
H_{ex}=-W\mathbb{I}_{2\times2}\otimes\left(\mathbf{n}\cdot\vec{\sigma}\right),\label{eq:exchange_field}
\end{equation}
with $\mathbf{n}=\left(\sin\theta_{F}\cos\varphi_{F},\sin\theta_{F}\sin\varphi_{F},\cos\theta_{F}\right)$
and $\vec{\sigma}=\left(\sigma_{1},\sigma_{2},\sigma_{3}\right)$,
which can be polarized in arbitrary direction $\mathbf{n}$. Choosing
$W=0.5\textrm{eV}$, states with one spin orientation reside at the
Fermi surface, while those with opposite orientation are shifted to
higher energies. This corresponds to a 100\% spin polarization of
the ferromagnet. A reduced spin polarization of the ferromagnet is
handled by a superposition of two calculations with opposite polarity
of the exchange field, as discussed in Appendix~\ref{sec:Analytical-TMR-ratio}.
The insulating barrier is modeled by a tunneling Hamiltonian of
the form
\begin{equation}
H_{T}=- C_B \sum_{k_x,k_y,\alpha.\sigma} d^\dagger_{k_x,k_y,\alpha.\sigma}
c_{k_x,k_y,\alpha.\sigma} + \mathrm{h.c.} \, ,\label{eq:thamilton}
\end{equation}
where $d^\dagger_{k_x,k_y,\alpha.\sigma}$ creates an electron in
orbital $\alpha$ with spin $\sigma$ in the top layer
of the topological insulator and $c_{k_x,k_y,\alpha.\sigma}$
destroys an electron in the bottom layer of the
ferromagnet. Here, $k_x$ and $k_y$ are the momentum
components perpendicular to the junction plane.
For the numerical calculations we have chosen a
small hopping matrix element of $C_{B}=0.1\textrm{eV}$, 
however, the relative TMR values calculated below
do not depend on this choice.

Due to the structure of the Hamiltonian Eq.~\eqref{eq:hamiltonian}
topological surface states appear on the topological insulator side
of the junction. The dispersion of the surface states forms a Dirac cone, 
which is hexagonally warped due to
the third order terms Eq.~(\ref{termr1}) and (\ref{termr2}). \cite{Fu2}
Note that the position of the Dirac point and the shape of the Dirac
cone depends on the surface direction. The surface states are spin-polarized
with the direction of the spin helically winding around the Dirac cone.
The spin-polarized tunneling current through the
barrier thus flows off into different directions in the topological insulator
depending on the polarization of the ferromagnet. In particular the
direction of the current in the topological insulator can be controlled
by rotation of the exchange field in the ferromagnet. This effect
can be used to construct a TMR device from such a junction.

\begin{figure}[t]
\centering \includegraphics[width=1\columnwidth]{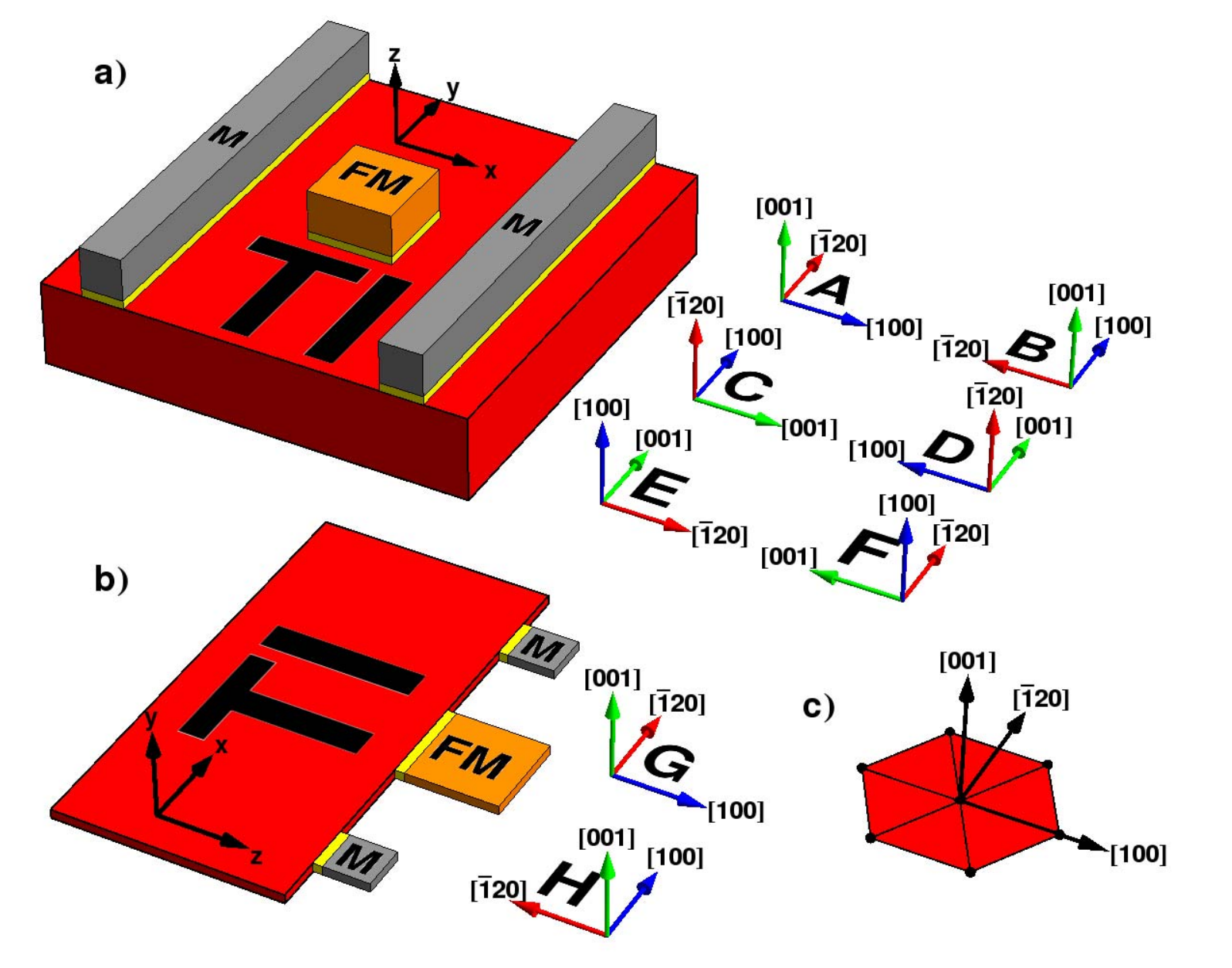} \caption{\label{fig2} 
Setups for TMR devices studied in this work. Different
orientations of the crystal axes of the TI relative to the electrodes
are given by the coordinate systems A-H, a) for a three-dimensional TI
and b) for a two-dimensional TI. In c) the crystallographic directions of 
the TI are shown.}
\end{figure}

Possible setups for such devices are shown in Fig.~\ref{fig2}. The
spin-polarized current that is injected via a ferromagnetic electrode
(FM) into the topological surface states can be extracted through
metallic electrodes (M) at opposite positions on the same surface.
For a given bias voltage the current through the two metallic electrodes
will differ in strength due to the spin-momentum locking of the surface
states. This leads to different resistances with respect to the ferromagnetic
electrode, because the spin direction of the tunneling electrons will
favor current flow to one of the two metallic electrodes. The resistances
will vary with the direction of the exchange field in the ferromagnet
resulting in a directional dependence of the resistance on the magnetization,
i.e. a TMR effect. 
Due to the anisotropy of Bi$_2$Se$_3$ we study six setups
for a three-dimensional (3D) device (A-F) and two for a two-dimensional
(2D) device (G, H) as shown in Fig.~\ref{fig2}, which may yield
different TMR ratios. These setups differ in the crystallographic
orientation of the topological insulator relative to the junction
plane. In the 3D cases we investigate the three clean
surfaces in $\left[001\right]$-, $\left[100\right]$- and $\left[\bar{1}20\right]$-direction
with two orthogonal arrangements of the metallic electrodes each.
The 2D case could be realized by a thin film strip contacted at the
edges, i.e. with an experimentally more complex arrangement. For that
purpose, the film thickness should be less than 6~nm (6 quintuple
layers) in order to ensure that the top and bottom surface states
become sufficiently hybridized. \cite{YiZhang} On the other hand
the film thickness should be at least 1.5~nm to avoid that the material
becomes topologically trivial. \cite{PD} Here, we consider a film
thickness of 3~nm to stay within these two limits. In this case the
parameter $M_{0}$ needs to be reduced to an effective 2D value of
$M_{0}=-0.197\textrm{eV}$, as has been discussed in Ref.~\onlinecite{PD}.
In the 2D case we do not consider thin films grown along $\left[100\right]$-
or $\left[\bar{1}20\right]$-direction as these are technically difficult
to realize. In the following, the cartesian coordinates $\left(x,y,z\right)$ are
tied to the geometry of the devices, while the orientation of the
crystal in these devices is expressed by Miller indices, as indicated
in Fig.~\ref{fig2}. Hence, $k_{z}$ is always the momentum perpendicular
to the surface plane and $k_{x}$ and $k_{y}$ are the in-plane momenta
with $k_{x}$ pointing from the junction towards the metallic electrode.
Note, that this definition of the cartesian coordinates only 
in the case of setup A coincides
with that in Eq.~\eqref{eq:hamiltonian} and \eqref{eq:exchange_field}.

The tunneling current through the insulating barrier is calculated
using Fermi's golden rule 
\begin{equation}
\Gamma_{mn}=\frac{2\pi}{\hbar}\delta\left(E_{n}-E_{m}\right)\left|\left\langle n\left|H_{T}\right|m\right\rangle \right|^{2},
\end{equation}
which gives the transition rate from an initial state $\left|m\right\rangle $
into a final state $\left|n\right\rangle $. Here, $H_{T}$ is the
tunneling Hamiltonian Eq.~\eqref{eq:thamilton} of the insulating barrier. \cite{Prange}
For a given bias voltage $U$ between the ferromagnetic electrode and the
topological insulator, the total
tunneling current is given by the expression \cite{Mahan} 
\begin{eqnarray}
I\left(U\right) & = & \frac{2\pi e}{\hbar}\sum_{m,n}\left[f\left(E_{m}-eU\right)-f\left(E_{n}\right)\right]\cdot\nonumber \\
 &  & \left|\left\langle n\left|H_{T}\right|m\right\rangle \right|^{2}\delta\left(E_{n}-E_{m}\right).
\end{eqnarray}
Here, the sum runs over all eigenstates $m$ and $n$ of the ferromagnet
and the topological insulator. The Fermi function 
\begin{equation}
f\left(E\right)=\frac{1}{1+e^{\frac{E}{k_{B}T}}}
\end{equation}
takes into account the occupation of the eigenstates at finite temperature.
Differentiating $I$ with respect to $U$ leads to the differential
conductance 
\begin{eqnarray}
G\left(U\right)=\frac{dI}{dU} & = & \frac{\pi e^{2}}{2\hbar k_{B}T}\sum_{m,n}\frac{1}{\cosh^{2}\frac{E_{m}-eU}{2k_{B}T}}\cdot\nonumber \\
 &  & \left|\left\langle n\left|H_{T}\right|m\right\rangle \right|^{2}\delta\left(E_{n}-E_{m}\right),\label{eq:dif_con}
\end{eqnarray}
which is used to define the TMR ratio 
\begin{equation}
\textrm{TMR}\left(U,\theta_{F},\varphi_{F}\right)=\frac{G_{\textrm{max}}\left(U\right)-G\left(U,\theta_{F},\varphi_{F}\right)}{G\left(U,\theta_{F},\varphi_{F}\right)}.\label{eq:TMR_ratio}
\end{equation}
Here, $\theta_{F}$ and $\varphi_{F}$ define the direction of the
exchange field Eq.~\eqref{eq:exchange_field} and $G_{\textrm{max}}\left(U\right)$
is the maximal differential conductance with respect to one metallic
electrode obtained when the exchange field and the polarization of
the surface electrons with propagation direction perpendicular to
that electrode are parallel. The maximal TMR ratio can then be obtained
by rotating the exchange field by $\pi$ into the antiparallel orientation
or, because of the spin-momentum locking of the TI surface states,
by comparing the differential conductances with respect to both metallic
electrodes (see Fig.~\ref{fig2}). To calculate the differential
conductance with respect to the different metallic electrodes, we
sum only over those eigenstates of the TI with a positive group velocity
component $v_{x}=\frac{1}{\hbar}\frac{\partial E}{\partial k_{x}}$
in the direction of the electrodes for one electrode and over those
with a negative group velocity component for the opposite electrode.

\section{Results\label{sec:results}}

\begin{figure}
\centering \includegraphics[width=0.95\columnwidth]{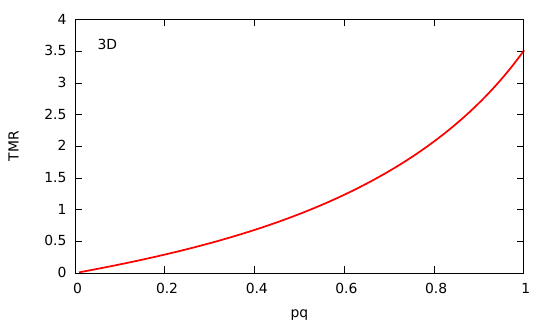} \includegraphics[width=0.95\columnwidth]{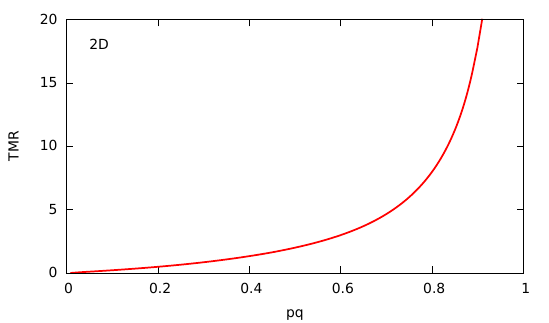}

\caption{\label{fig:TMR_pol} Maximal TMR ratio as a function of the product
of the surface state polarization $p$ and the polarization $q$ of
the ferromagnet within the analytical approximations Eq.~(\ref{eq:TMR_3D})
and Eq.~(\ref{eq:TMR_2D}) }
\end{figure}

In this section we present our results for the TMR ratio in Bi$_{2}$Se$_{3}$
based on numerical calculations as well as an analytical approximation.
The analytical approximation is derived in Appendix~\ref{sec:Analytical-TMR-ratio}.
It assumes isotropic helical surface states and neglects the influence
of bulk states. This should be a good approximation for bias voltages
chosen within the bulk gap of the material in the vicinity of the Dirac point.
While the numerical calculations give an exact solution based on Hamiltonian
Eq.~\eqref{eq:hamiltonian} including the influence of the bulk states,
the analytical approximation gives a formula depending on the 
spin-polarization $p$ of the topological surface states. Here, $0\le p\le1$
is the average magnitude of the spin expectation value in the vicinity of
the Dirac point. The analytical formula is useful as it provides a
quick estimate of the TMR ratio and can be used for different materials
once the spin-polarization $p$ of the surface states is known. Depending
on the dimensionality of the device the formulas for the TMR ratio
are slightly different, but independent of temperature and bias voltage
(see Appendix~\ref{sec:Analytical-TMR-ratio}): 
\begin{equation}
\textrm{TMR}_{\textrm{3D}}\left(p,q,\varphi_{F}\right)=\frac{1+\sin\varphi_{F}}{\frac{\pi}{2pq}-\sin\varphi_{F}}\label{eq:TMR_3D}
\end{equation}
\begin{equation}
\textrm{TMR}_{\textrm{2D}}\left(p,q,\varphi_{F}\right)=\frac{1+\sin\varphi_{F}}{\frac{1}{pq}-\sin\varphi_{F}}.\label{eq:TMR_2D}
\end{equation}
Here, $0\le q\le1$ is the spin-polarization of the ferromagnet. 
The angle $\varphi_{F}$ is
the in-plane polarization angle of the exchange field. The maximal
TMR ratio is reached after a $\pi$ rotation of the exchange field
($\varphi_{F}=\frac{\pi}{2}$) and is shown as a function of $pq$
in Fig.~\ref{fig:TMR_pol}. It is limited to $\textrm{TMR}_{\textrm{3D}}\left(1,1,\frac{\pi}{2}\right)\approx3.5$
for a 3D device while it diverges in the 2D case for $pq\rightarrow1$.
The factor $\pi/2$ in the denominator of Eq.~(\ref{eq:TMR_3D})
as compared with Eq.~(\ref{eq:TMR_2D}) comes from the fact that
the two-dimensional manifold of surface states on the 3D devices possesses
in-plane spin polarizations of all directions, as shown in Appendix~\ref{sec:Analytical-TMR-ratio}.
Even for surface states with $p=1$ a finite amount of electrons
can thus tunnel into surface states with a velocity component into
the opposite direction, reducing the TMR ratio. In the 2D devices
such a situation can be avoided.

For the numerical TMR ratios we calculate the eigenstates and eigenenergies
of the TI by an exact diagonalization of the Fourier-transformed Hamiltonian
on a lattice of size $50\times400\times400$ for setups A and B, $100\times400\times400$
for C-F, and $200\times800$ for G and H, where the first number is
always for the direction perpendicular to the surface plane. Due to
the periodic boundary conditions within the surface plane this can
be done separately for all discrete in-plane momenta $\tilde{\mathbf{k}}$.
To obtain the sign of the group velocity component $v_x$ we compare
the eigenenergies with those we get after a small variation of $k_x$.
Assuming the FM to be large, the spatial
dependence of the FM states perpendicular to the surface is given
by $\sin zk_{z}$, where $z$ is the lattice position and the perpendicular
momentum $k_{z}$ is a continuous function of the in-plane momentum
$\tilde{\mathbf{k}}$ and the TI eigenenergies, satisfying energy
and in-plane momentum conservation (see
Appendix~\ref{sec:Analytical-TMR-ratio}). To obtain
$G_{\textrm{max}}\left(U\right)$ the direction of the 
magnetization of the ferromagnet is chosen such that the differential
conductance is maximized. This occurs, when the magnetization
direction fits the expectation value of the spin
operators $\Sigma_{i}=\mathbb{I}_{2\times2}\otimes\sigma_{i}$ \cite{Silvestrov}
of those TI surface states that propagate
towards a metallic electrode. The
TMR ratio is then calculated using Eq.~\eqref{eq:TMR_ratio},
where the maximal and minimal differential conductances are calculated
by summing only over those states with positive or negative group
velocity component $v_x$ in Eq.~\eqref{eq:dif_con}, i.e. by looking at opposite
metallic electrodes.

\begin{figure}
\centering \includegraphics[width=0.95\columnwidth]{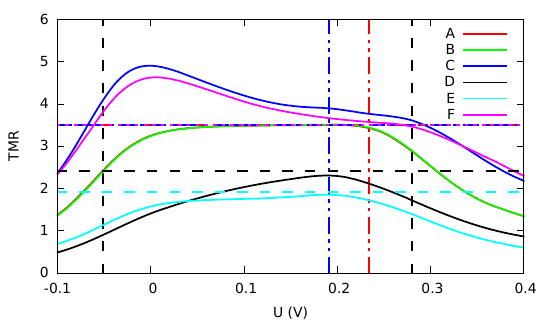} \includegraphics[width=0.95\columnwidth]{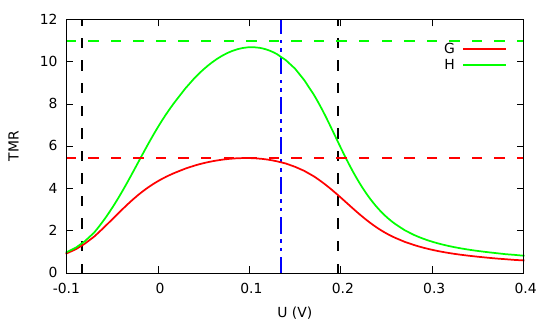}

\caption{\label{fig:numerical_results} TMR ratio based on numerical calculations
for the different setups A-G in Fig.~\ref{fig2}. The horizontal dashed lines
represent the analytical approximation for the spin polarization $p$
of the surface states close to the Dirac point. The energy position
of the Dirac point is shown by the vertical red dashed-dotted line for the
[001]-surface (234 meV)
and the vertical blue dashed-dotted line for the [100] and
$\left[\bar{1}20\right]$ surfaces
(191 meV in the 3D case and 135 meV in the 2D case). 
The vertical black
dashed lines denote the positions of the gap edges.}
\end{figure}

In Fig.~\ref{fig:numerical_results} the results of these calculations
at room temperature are shown (solid lines) in comparison with the
corresponding expectation from the analytical approximation 
(horizontal dashed lines with the same color).
The FM is chosen to be ideal for these calculations, i.e. fully polarized
($q=1$). To make an appropriate comparison of our numerical results with the 
analytical approximation we have determined the
polarization $p$ of the TI surface states from the numerical calculation
in the following way: we determine $p$ from the absolute value
of the spin expectation value for a surface state having a momentum
in the direction of the metallic electrode close to the Dirac point.
As the magnetization
of the ferromagnet is aligned to the spin of this state, the analytical formula
gives the best approximation for this value of $p$, because the main contribution
to the differential conductance comes from electrons with similar
spin direction. Note, that
the value of $p$ depends on the crystallographic orientation of the surface
leading to different values for the different setups.

\begin{figure}
\centering \includegraphics[width=0.95\columnwidth]{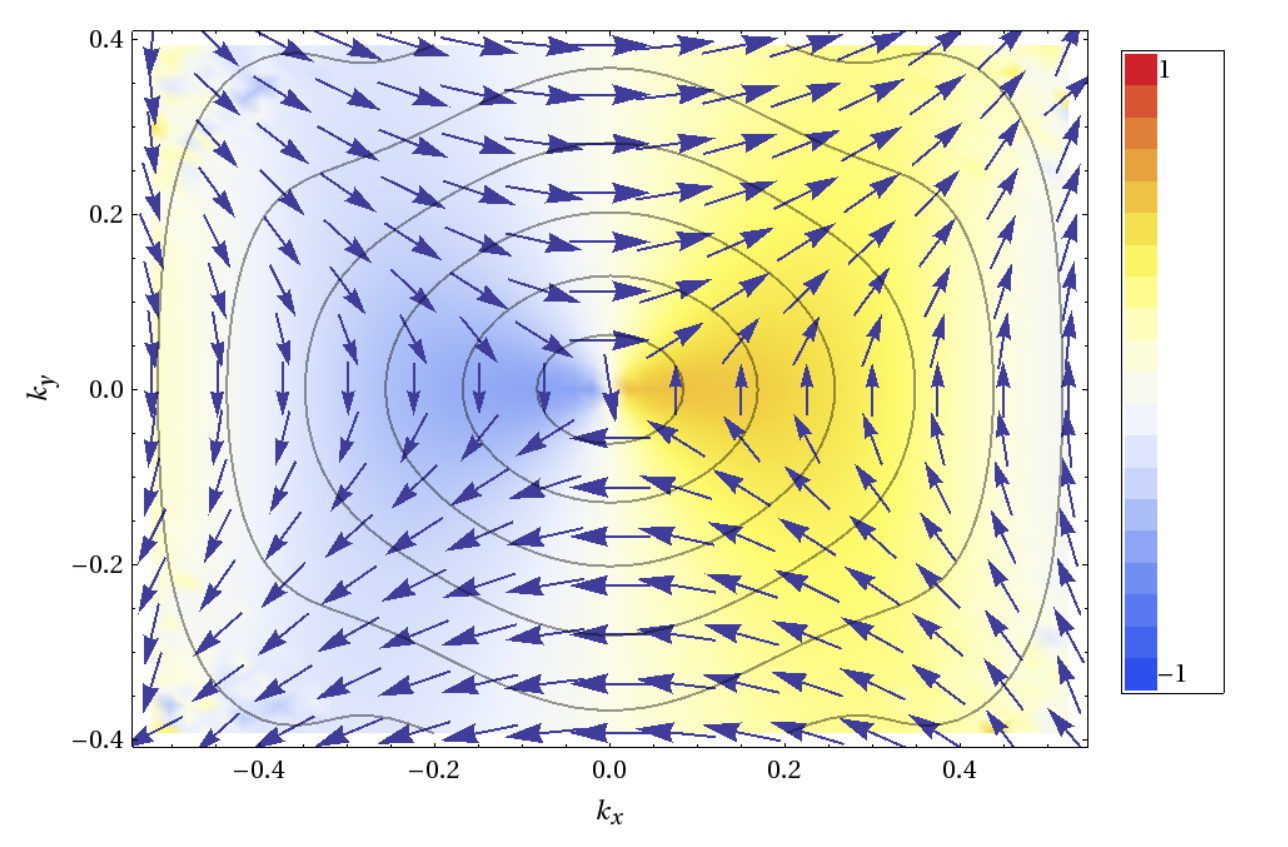} \includegraphics[width=0.95\columnwidth]{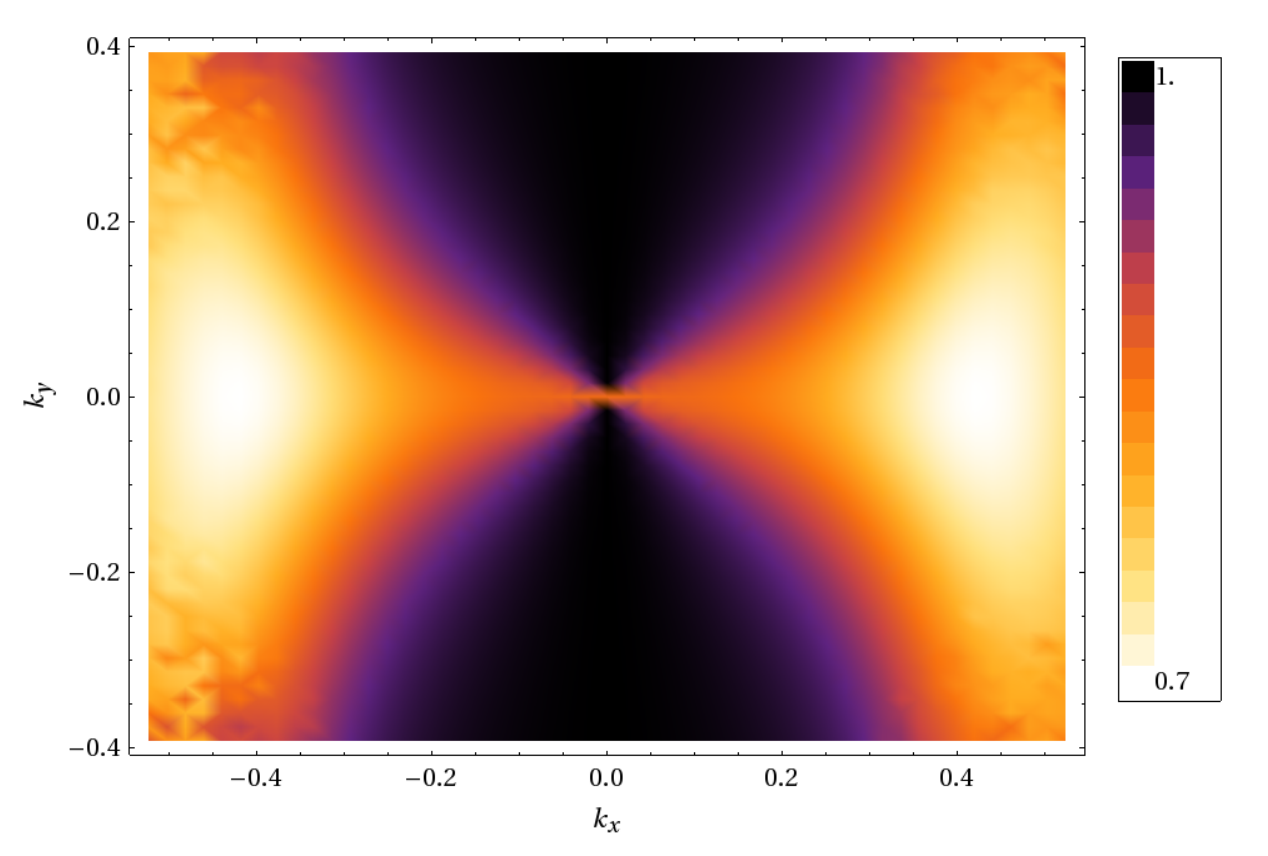}

\caption{\label{fig:Spin-expectation-values} Spin expectation values of the
lower Dirac cone on the side surface in $\left[\bar{1}20\right]$-direction.
The declaration of the momentum components corresponds to setup D.
In the upper figure the in-plane spin is shown by the arrows while
the out-of-plane spin component is indicated by the color. The lines
are constant energy contours. The lower figure shows the absolute
value of the spin expectation value.}
\end{figure}

Since the surface states only exist inside the bulk gap and the bulk
states are unpolarized, the TMR ratio drops towards the edges of the
gap (shown by the vertical black lines). This influence of the bulk
states increases with increasing temperature and may reduce the maximal
TMR ratio in systems with small gaps. Taking a closer look at the
different setups, setup A and B show nearly no difference and are
in very good agreement with the analytical approximation $\textrm{TMR}_{\text{3D}}\approx3.5$
for $p=1$. There is only a tiny deviation for larger $\tilde{\mathbf{k}}$,
where the numerical TMR ratio decreases due to a small albeit increasing
out-of-plane polarization of the surface states caused by hexagonal
warping.

Because of the anisotropy of the $\left[100\right]$ and
$\left[\bar{1}20\right]$ surfaces, the analytical approximation
gives only a rough estimate for the TMR ratio of setups C-F. The anisotropy
changes the shape of the Dirac cone and with it the spin orientation
from circular to a more elliptical form (see Fig.~\ref{fig:Spin-expectation-values}).
This creates an imbalance in the tunneling probabilities of electrons
with different spin orientations. For setups C and F the TMR ratio
at the Dirac point is slightly larger than the estimated 3.5 and rises
up to $\sim4.9$ for setup C and $\sim4.6$ for setup F near the lower
edge of the bulk gap. As the TMR ratio increases away from the Dirac
point, it is limited by the size of the bulk gap and the position
of the Dirac point inside the gap. The fact that setup C reaches a
larger value than setup F indicates that for equally polarized surfaces
it is beneficial to choose the surface parallel to a crystal axis.

In contrast to setups C and F, the anisotropy reduces the TMR ratio
for setups D and E, which is already smaller because the surface states
have a reduced spin-polarization of $p\approx0.86$ and $p\approx0.77$
near the Dirac point, respectively. The TMR ratio is further reduced
by a change in the polarization orientation and strength as a function
of energy away from the Dirac point.

In the case of a 2D device (setup G and H), the TMR ratio reaches
much larger values than in the 3D case, as there are basically only
two spin orientations flipping at $\tilde{\mathbf{k}}=0$. Near the
Dirac point, the TMR ratio now falls somewhat below the analytical
approximation and decreases rapidly towards the edges of the gap.
On one hand this is due to the higher sensitivity of the larger TMR
ratios resulting in a more important influence of the bulk states.
On the other hand, similar as for setup D and E, there is a change
in the orientation and strength of the spin polarization as a function
of energy, further reducing the TMR ratio away from the Dirac point.
Still we get a maximal TMR ratio of $\sim5.4$ for setup G ($p\approx0.73$)
and $\sim10.7$ for setup H ($p\approx0.85$) at room temperature.

\begin{figure}
\centering\includegraphics[width=0.95\columnwidth]{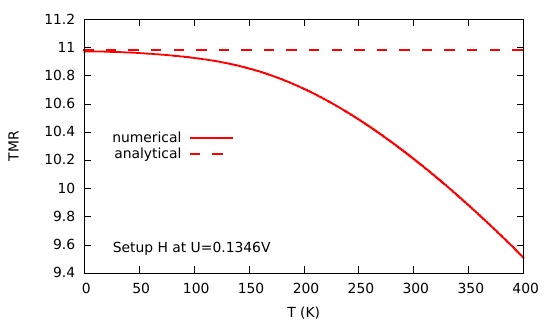}

\caption{\label{fig:TMR_temp} Temperature dependence of the TMR ratio for
the 2D setup H at the Dirac point (solid line). The dashed line
shows the TMR ratio within the analytical approximation, which
is reached by the numerical results at low temperature. }
\end{figure}

In Fig.~\ref{fig:TMR_temp} we show the temperature dependence of
the TMR ratio for setup H and bias voltage chosen at the Dirac point.
Here, it is seen that the value of the analytical approximation is
reached at low temperatures. This result demonstrates that the finite
temperature occupation of the bulk states somewhat mitigates the TMR
ratio already at room temperature in the 2D devices.

\begin{figure}
\centering\includegraphics[width=0.9\columnwidth]{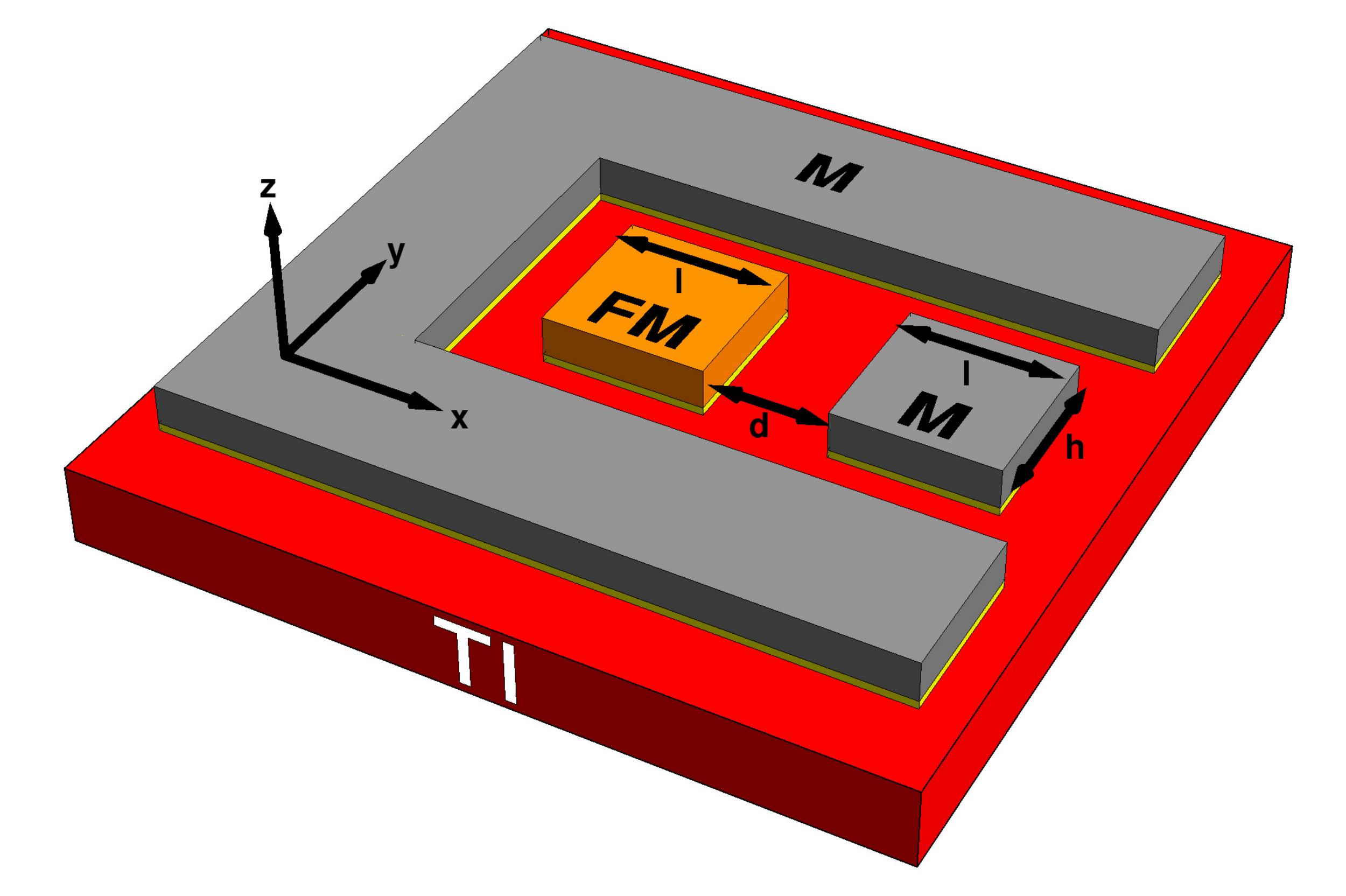}

\caption{\label{fig:2Dapprox} Alternative device on the surface
of a 3D TI which allows to approach the large TMR ratios of a 2D device. 
The U-shaped metallic electrode captures nearly all electrons
moving in other directions than the positive x-direction.
The ferromagnetic electrode and the smaller metallic one possess
length $l$ and height $h$. The distance between them is $d$.}
\end{figure}

\begin{figure}
\centering\includegraphics[width=0.9\columnwidth]{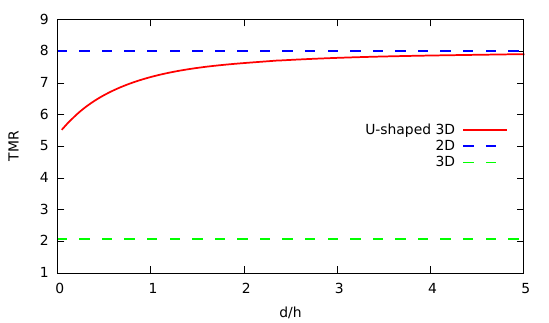}

\caption{\label{fig:TMR-ratio-2Da}TMR ratio near the Dirac point
for the device shown in Fig.~\ref{fig:2Dapprox}
as a function of $\frac{d}{h}$ for $l=h$ and $pq=0.8$. A higher
ratio $\frac{l}{h}$ further increases the TMR-ratio. The dashed lines
show the results of the 2D and 3D devices discussed above.}

\end{figure}
The 2D devices (setup G and H) reach much larger TMR values than
the 3D ones. However, 
they are technically more challenging to realize as the tunneling
barrier has to be attached to the edge states of a thin TI film.
In Fig.~\ref{fig:2Dapprox} we propose an alternative 3D device, which
is easier to realize than the 2D devices, however, approaches the
same large TMR values. In this device one of the two metallic electrodes
is U-shaped and is supposed to capture most electrons that do not
propagate in positive $x$-direction from the ferromagnetic to the
other, smaller metallic electrode. As a result, the electrons that reach
the smaller metallic electrode preferentially possess only
a single spin orientation. Thus, the transport between the
ferromagnetic and the smaller metallic electrode approaches
a one-dimensional transport, similarly as in the edge states
of the 2D devices. We have applied our analytical approximation 
to this geometry (for more details see Appendix~\ref{sec:Analytical-TMR-ratio}),
corresponding to a $[001]$ surface, i.e. the crystallographic
orientation of setup A.
In Fig.~\ref{fig:TMR-ratio-2Da} the TMR ratio of this U-shaped
device is shown as function of the ratio $d/h$, where $d$ is
the distance between the ferromagnetic and the smaller metallic
electrode, and $h$ is the height of the electrodes as shown
in Fig.~\ref{fig:2Dapprox}. In this calculation we assumed that
the ferromagnetic and the smaller metallic electrode are
square-shaped, i.e. $l=h$. For illustration, we assumed
that the product $pq$ of the polarizations of the ferromagnet
and the topological insulator is 0.8. As Fig.~\ref{fig:TMR-ratio-2Da}
shows, the U-shaped 3D device yields much larger TMR values
than the 3D devices. For $d/h \gg 1$ the TMR ratio of the U-shaped device
approaches the TMR ratio of the 2D device, which is clear
as with increasing distance $d$ the angular range of
topological surface states that can reach the metallic electrode
is gradually reduced and focused.

The calculations we reported here were done with parameters appropriate
for undoped Bi$_2$Se$_3$, as pointed out above. However, in practice
these materials often appear to be intrisically doped, \cite{Xia,Hsieh2}
which shifts the Fermi level with respect to the surface Dirac cone.
For Bi$_2$Se$_3$ a Fermi level shift of 200~meV has been reported. \cite{Xia} 
We want to point out that the functionality of the devices proposed here
is not affected by such intrinsic doping. The Fermi level shift will
shift the energies $E_n$ of the topological surface states with respect
to the ones of the ferromagnet. This energy shift can be fully compensated
for by a shift of the bias voltage $U$ in Eq.~\eqref{eq:dif_con}.
This will lead to a bias voltage shift of our results in 
Fig.~\ref{fig:numerical_results}. Thus the high TMR ratios will
be robust against intrinsic doping, but just appear at a shifted
bias voltage.

While our calculations find a full spin-polarization for the z-surface,
consistent with recent spin- and angle-resolved photoemission spectroscopy
(SARPES) measurements of Pan et al\cite{Pan:PRB88}, the spin-polarization
in Bi$_{2}$Se$_{3}$ is still controversial. The reported spin-polarization
ranges from 50\%-65\% in first principle calculations \cite{Yazyev,Wang,Sanchez}
to 75\%-80\% or more in SARPES measurements \cite{Pan:PRL106,Jozwiak}.
If we consider these values, according to Fig.~\ref{fig:TMR_pol}
the TMR ratio could be reduced to $\sim0.9-2$ in the 3D devices and
$\sim2-8$ in the 2D devices, which is still large, though.

As our analytical formulas for the TMR ratio Eq. \eqref{eq:TMR_3D}
and \eqref{eq:TMR_2D} depend only on the spin-polarization and not
on the exact parametrization they are universally valid for all materials
in the limit of an isotropic surface. In contrast, the TMR ratio for
an anisotropic surface strongly depends on the model parameters and
the deviation from the isotropic solution may therefore be different
for other materials.

In the present work we are focussing on Bi$_{2}$Se$_{3}$, which
has been studied well in the past. Considering other materials, TlBiSe$_{2}$
could also be a good candidate for TMR devices as discussed here,
because it has an in-plane spin-polarization of $\sim80$\%, a negligible
out-of-plane polarization and a large bulk gap of $\sim0.35$eV.
\cite{Souma:PRL109,Souma:PRL106,Sato}
The high spin polarization of 80\%-90\% of Sb$_{2}$Te$_{3}$ promises
a high TMR ratio, too. However, at room temperature it will probably
be reduced because of the small bulk gap of only $\sim0.2\textrm{eV}$.\cite{Pauly}
Bi$_{2}$Te$_{3}$ and Pb(Bi,Sb)$_{2}$Te$_{4}$ seem to be less attractive
since they have in-plane spin-polarizations of only 45\%-60\%
\cite{Yazyev,Herdt,Souma:PRL106,Scholz} and 50\%\cite{Nomura} and small 
bulk gaps of only $\sim0.08-0.165\textrm{eV}$\cite{Yazyev,Chen}
and $\sim0.2-0.23\textrm{eV}$\cite{Souma:PRL108,Kuroda}, respectively.
They also possess a significant out-of-plane polarization due to the
hexagonal deformation of the Fermi surface. \cite{Yazyev,Souma:PRL106,Herdt,Nomura}

An attractive feature of the devices proposed here is the
topological protection of the spin-locked surface states,
which makes them particularly robust against perturbations.
In usual TMR junctions defects close to the interface can
substantially suppress the TMR ratio in particular at high temperature.
\cite{Yamamoto,Drewello} For the present devices we have to distinguish
non-magnetic and magnetic scattering mechanisms.
Non-magnetic scattering processes, like disorder or faceting
of the barrier, which scatter the momentum of an electron
during the transfer from the ferromagnet to the topological insulator, 
but keep its spin conserved, are only weakly 
affecting our TMR results. This is because the TMR ratio is
predominantly dictated by the helical spin structure of the
surface states, which is topologically protected.
As long as the spin is conserved, momentum scattering is
not changing our results as the total current already consists of
contributions from all momentum directions.
Magnetic scattering mechanisms, on the other hand, do
reduce the TMR ratio. This is obvious, as a spin-flip
process during tunneling will change the direction of
flow in the topological surface state. Such processes
could be modeled by a reduced apparent polarization $q$
of the ferromagnet seen by the topological insulator.
An experimental realization of our devices should thus
take care to avoid magnetic scattering in the barrier.
These are the same quality requirements as for conventional 
TMR devices based on magnetic tunnel junctions and has been 
successfully dealt with in the past, however.

The distance between the electrodes should be chosen
sufficiently small such that electrons travelling from the
ferromagnetic electrode to one metallic electrode are not
back-scattered to the opposite metallic electrode inbetween.
As the surface states are topologically protected by
time-reversal symmetry, such back-scattering is
strongly suppressed and can only occur when time-reversal 
symmetry breaking scattering impurities or imperfections exist 
in the topological insulator. Thus, the distance between the electrodes 
should be smaller than the spin-flip mean-free-path of the surface states.
Values of the order of 2~$\mu$m have been reported. \cite{Koenig1,Koenig2}
It has been demonstrated in the past that magnetic tunnel junctions
can be structured down to a 50 nanometer scale. \cite{Kubota}

In conventional semiconductor Rashba devices
TMR ratios of up to 50\% have been reached at room temperature. \cite{Wojcik,Yoh}
For TMR devices based on magnetic tunnel junctions with two ferromagnetic
electrodes reported TMR ratios range between 180\% and 600\%. 
\cite{Yamamoto,Drewello,Parkin,Yuasa,Ikeda,Liu}
The TMR values calculated here are comparable or larger than these values.

\section{Summary and Conclusions}

In the present work we proposed specific setups to build TMR devices
exploiting the topological insulator material Bi$_{2}$Se$_{3}$.
Using a realistic tight-binding model we calculated the achievable
TMR ratios as a function of bias voltage and temperature. If the bias
voltage is chosen properly, we find that room temperature
TMR ratios of 490~\% can be achieved in the 3D devices and 1070~\% 
in the 2D devices. We suggested a U-shaped 3D device, which is
experimentally easier to realize than the 2D devices, but reaches
correspondingly high TMR values. We derived simple analytical formulas for the 
devices, which provide a quick approximate figure of merit once
the surface state polarization of a given topological insulator is
known. For the 3D devices we have shown that on an anisotropic surface
the TMR ratio may become larger than the one estimated from the analytical
formula. In contrast to conventional TMR devices the devices proposed
here do not need a second ferromagnetic layer, as an intrinsic magnetic
reference direction is already provided by the topological insulator.
In addition, the present devices provide a means to experimentally
probe the surface state polarization of a given topological insulator.


\appendix

\section{Analytical TMR Ratio\label{sec:Analytical-TMR-ratio}}

In this appendix we derive an analytical approximation for the TMR
ratio. We base this on an analytical approximation of the surface
states only and neglect the effects of bulk states. Considering the
large bulk gap of Bi$_{2}$Se$_{3}$, this should be a good approximation
even at room temperature.

To find an analytical approximation of the surface states for a surface
at $z=0$, we expand Hamiltonian Eq.~\eqref{eq:hamiltonian} up to
second order in $k_{z}$. As a starting point we first neglect all
$k_{x}$- and $k_{y}$-dependent terms by setting $k_{x}=k_{y}=0$.
If we assume the TI to be half infinite with its boundary at $z=0$,
we have to replace $k_{z}$ with the momentum operator $-i\partial_{z}$
and search for nontrivial solutions that vanish both at $z=0$ and
for $z\rightarrow\infty$. There exist two degenerate eigenstates
with energy $E=-\frac{C_{1}M_{0}}{M_{1}}$ satisfying both conditions:
\begin{eqnarray}
\psi_{1}\left(z\right) & = & \frac{1}{N}\left(\begin{array}{c}
\sqrt{\frac{M_{1}-C_{1}}{2M_{1}}}\\
0\\
\sqrt{\frac{M_{1}+C_{1}}{2M_{1}}}\\
0
\end{array}\right)\left(e^{-\alpha_{1}z}-e^{-\alpha_{2}z}\right)\\
\psi_{2}\left(z\right) & = & \frac{1}{N}\left(\begin{array}{c}
0\\
\sqrt{\frac{M_{1}-C_{1}}{2M_{1}}}\\
0\\
\sqrt{\frac{M_{1}+C_{1}}{2M_{1}}}
\end{array}\right)\left(e^{-\alpha_{1}z}-e^{-\alpha_{2}z}\right),
\end{eqnarray}
with $\alpha_{1}\neq\alpha_{2}$, $\Re\alpha_{1/2}>0$ and some normalization
constant $N$ for the $z$-dependent part. Next we treat the neglected
$k_{x}$- and $k_{y}$-dependent terms as a perturbation using degenerate
perturbation theory and get the new surface states 
\begin{equation}
\psi_{\pm}\left(k_{x},k_{y},z\right)=u_{1}\psi_{1}+u_{2}\psi_{2},
\end{equation}
with 
\begin{eqnarray}
\lefteqn{u_{1}\left(k_{x},k_{y}\right)=}\\
 &  & -\frac{\left(m_{1}-im_{2}\right){\mathrm{sgn}}\left(\mathcal{R}_{1}\mp\sqrt{m_{1}^{2}+m_{2}^{2}+\mathcal{R}_{1}^{2}}\right)}{\sqrt{m_{1}^{2}+m_{2}^{2}+\left(\mathcal{R}_{1}\mp\sqrt{m_{1}^{2}+m_{2}^{2}+\mathcal{R}_{1}^{2}}\right)^{2}}}\nonumber \\
\lefteqn{u_{2}\left(k_{x},k_{y}\right)=}\\
 &  & \frac{\sqrt{\left(\mathcal{R}_{1}\mp\sqrt{m_{1}^{2}+m_{2}^{2}+\mathcal{R}_{1}^{2}}\right)^{2}}}{\sqrt{m_{1}^{2}+m_{2}^{2}+\left(\mathcal{R}_{1}\mp\sqrt{m_{1}^{2}+m_{2}^{2}+\mathcal{R}_{1}^{2}}\right)^{2}}}.\nonumber 
\end{eqnarray}
The perturbed eigenenergies are 
\begin{equation}
E_{\pm}=-\frac{C_{1}M_{0}}{M_{1}}+\epsilon_{0}^{\prime}-\frac{C_{1}}{M_{1}}m_{4}^{\prime}\pm\sqrt{\left(m_{1}^{2}+m_{2}^{2}+\mathcal{R}_{1}^{2}\right)\left(1-\frac{C_{1}^{2}}{M_{1}^{2}}\right)}.
\end{equation}
Here, $\epsilon_{0}^{\prime}=\frac{4}{3}C_{2}\left(3-2\cos\frac{1}{2}k_{x}\cos\frac{\sqrt{3}}{2}k_{y}-\cos k_{x}\right)$
and $m_{4}^{\prime}=\frac{4}{3}M_{2}\left(3-2\cos\frac{1}{2}k_{x}\cos\frac{\sqrt{3}}{2}k_{y}-\cos k_{x}\right)$
and $\pm$ is for the upper and lower Dirac cone, respectively. If
we expand $u_{1}$, $u_{2}$ and $E_{\pm}$ up to second order in
$k_{x}$ und $k_{y}$, the solutions become isotropic in the $k_{x}$-$k_{y}$-plane
and we can write $\psi_{\pm}$ solely as a function of $z$ and the
in-plane polar angle $\varphi$. The energies $E_{\pm}$ then become
\begin{equation}
E_{\pm}=-\frac{C_{1}M_{0}}{M_{1}}+\left(C_{2}-\frac{C_{1}}{M_{1}}M_{2}\right)k^{2}\pm A_{0}\sqrt{1-\frac{C_{1}^{2}}{M_{1}^{2}}}k
\end{equation}
and thus only depend on the magnitude $k=\sqrt{k_{x}^{2}+k_{y}^{2}}$.
This expression agrees with the one derived previously in
Ref.~\onlinecite{Shan}.

Calculation of the expectation values of the spin operators $\Sigma_{i}=\mathbb{I}_{2\times2}\otimes\sigma_{i}$
\cite{Silvestrov} shows that the spin of these surface states is
always perpendicular to the in-plane momentum and lies within the
surface plane. As the direction of the spin is the same for both orbitals,
this model yields surface states which are fully polarized, at least
at the $[001]$-surface. Here, we allow for a finite spin polarization
$0\le p\le1$ and thus rewrite the surface state wave functions as
a function of polarization $p$ 
\begin{equation}
\psi_{\pm}\left(p,\varphi\right)=\frac{1}{2}\left(\begin{array}{c}
\pm\sqrt{1+p}e^{-i\left(\varphi-\frac{\pi}{2}\right)}\\
\sqrt{1+p}\\
\mp\sqrt{1-p}e^{-i\left(\varphi-\frac{\pi}{2}\right)}\\
\sqrt{1-p}
\end{array}\right).
\end{equation}
As all constant prefactors cancel in the TMR ratio we neglect these
as well as the spatial dependence, because in the calculation of the
tunneling current only the lattice position next to the barrier plays
a role. The spatial dependence of the FM states perpendicular to the 
surface is a superposition of an incoming plane wave $e^{i z k_z}$ 
with the reflected one $e^{-i z k_z}$ having a node inside the barrier 
(at $z=0$) and is thus given by $\sin z k_{z}$.
Note that this is an exact solution of the tight-binding model
for a thick FM with zero transmission probability through the barrier.
Here, the momentum $k_{z}$ is a continuous function of the in-plane
momentum $k$ and the TI eigenenergies $E_{\pm}$, satisfying energy
and in-plane momentum conservation: 
\begin{equation}
k_{z}(k)=\arccos\frac{C_{F}\left(2+k^{2}\right)\pm W-E_{\pm}}{2C_{F}}
\end{equation}
Since the spin of the surface states lies within the surface plane
it is sufficient to write the states of the FM as a function of the
in-plane polar angle $\varphi_{F}$ 
\begin{equation}
\psi_{F}\left(k,\varphi_{F}\right)=\frac{1}{2}\sin k_{z}(k)\left(e^{-i\varphi_{F}},1,e^{-i\varphi_{F}},1\right).
\end{equation}
The transfer matrix elements 
\begin{equation}
\left|\left\langle \psi_{F}\left|H_{T}\right|\psi_{\pm}\right\rangle \right|^{2}=C_{B}^{2}\sin^{2}k_{z}(k)\left[1\mp p\sin\left(\varphi_{F}-\varphi\right)\right]
\end{equation}
can then be inserted into Eq.~\eqref{eq:dif_con}.

In the calculation of the differential conductance the geometry
of the device can be accounted for by an additional angular dependent
factor $f\left(\varphi\right)$. This factor describes the probability that an electron
starting its propagation in the TI at an angle $\varphi$ below the FM
electrode ends up
at the metallic electrode at $x>0$ averaged over the spatial extend of the
two electrodes. Under the assumption that
$f\left(\varphi\right)=f\left(-\varphi\right)$,
which holds for a device which is mirror symmetric with respect to the
$x$-$z$-plane and valid for all devices considered in this work,
the $\varphi$-integral can be separated from the rest of the differential
conductance 
\begin{eqnarray}
\lefteqn{G\left(T,U,\varphi_{F}\right)=}\\
 &  & \frac{\textrm{const.}}{T}\int_{0}^{k_{0}}dkk\Bigg(\int_{-\pi}^{\pi}d\varphi f\left(\varphi\right)\frac{\left|\left\langle \psi_{F}\left|H_{T}\right|\psi_{+}\right\rangle \right|^{2}}{\cosh^{2}\left(\frac{E_{+}-eU}{2k_{B}T}\right)}+\nonumber \\
 &  & f\left(\varphi-\pi\right)\frac{\left|\left\langle \psi_{F}\left|H_{T}\right|\psi_{-}\right\rangle \right|^{2}}{\cosh^{2}\left(\frac{E_{-}-eU}{2k_{B}T}\right)}\Bigg)\nonumber \\
 & = &
\int_{-\pi}^{\pi}d\varphi f\left(\varphi\right) \left(1-p\sin\varphi_{F} \cos\varphi\right)G'\left(T,U\right), \nonumber
\end{eqnarray}
with the function
\begin{eqnarray}
G'\left(T,U\right) & = & \frac{\textrm{const.}}{T}\int_{0}^{k_{0}}dkk\cdot\\
 &  & \left(\frac{\sin^2 k_{z}(k)}{\cosh^{2}\left(\frac{E_{+}-eU}{2k_{B}T}\right)}+\frac{\sin^2 k_{z}(k)}{\cosh^{2}\left(\frac{E_{-}-eU}{2k_{B}T}\right)}\right)\nonumber 
\end{eqnarray}
which is independent of $\varphi_{F}$. 

Up to this point we considered an ideal FM, i.e. fully polarized. To account
for a finite polarization $q=n_{+}-n_{-}$ of the FM we have to replace
$G\left(T,U,\varphi_{F}\right)$ by 
\begin{eqnarray}
\lefteqn{G''\left(T,U,\varphi_{F}\right)=}\\
 &  & n_{+}G\left(T,U,\varphi_{F}\right)+n_{-}G\left(T,U,-\varphi_{F}\right)\nonumber \\
 & = & \int_{-\pi}^{\pi}d\varphi f\left(\varphi\right)
\left(1-pq\sin\varphi_{F} \cos\varphi\right)G'\left(T,U\right).\nonumber
\end{eqnarray}
Here, $0\leq n_{\pm}\leq1$ with $n_{+}+n_{-}=1$ is the relative
density of states of electrons in the FM with spin projection parallel
(+) or antiparallel (-) to the polarization angle $\varphi_{F}$.
Defining 
\begin{equation}
\gamma=\frac{\int_{-\pi}^{\pi}d\varphi f\left(\varphi\right)}{\int_{-\pi}^{\pi}d\varphi f\left(\varphi\right)\cos\varphi},
\end{equation}
the TMR ratio 
\begin{equation}
\textrm{TMR}\left(p,q,\varphi_{F}\right)=\frac{1+\sin\varphi_{F}}{\frac{\gamma}{pq}-\sin\varphi_{F}}
\end{equation}
only depends on $p$, $q$ and $\varphi_{F}$. In particular it does
not depend on temperature, because the temperature dependence of the
differential conductance $G'(T,U)$ is the same in all directions
due to the in-plane rotational symmetry of the surface states and
thus drops out in the TMR ratio. This temperature independence is
lost of course, if the contribution of the bulk states is included.
The geometry of the device only appears via the single parameter $\gamma$.
This parameter can now be calculated for different
geometries of the devices. For setups A-H we assume that all electrons 
initially moving in positive $x$-direction end up at the electrode at 
$x>0$ and all others at the counterelectrode. 
For the 3D devices A-F the function $f\left(\varphi\right)$
is then simply given by
\begin{equation}
f\left(\varphi\right)=\begin{cases}
1 & \textrm{if }\varphi \in \left[-\frac{\pi}{2},\frac{\pi}{2}\right]\\
0 & \textrm{else}
\end{cases}
\end{equation}
which yields $\gamma=\frac{\pi}{2}$. For the 2D devices G and H we have 
\begin{equation}
f\left(\varphi\right)=\delta\left(\varphi\right)
\end{equation}
giving $\gamma=1$.
\begin{figure}[t]
\centering\includegraphics[width=0.9\columnwidth]{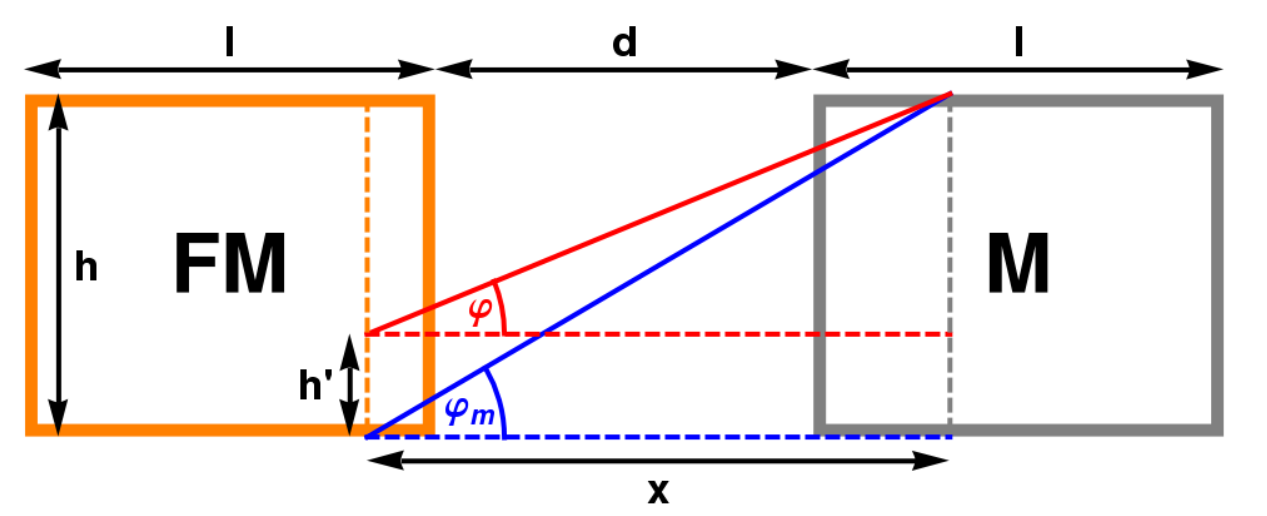}

\caption{\label{fig:Sketch} Sketch of the electrodes in the device 
from Fig.~\ref{fig:2Dapprox}, illustrating the calculation
of the angular probability distribution $f\left(\varphi\right)$.}

\end{figure}
\begin{figure}[t]
\centering\includegraphics[width=0.9\columnwidth]{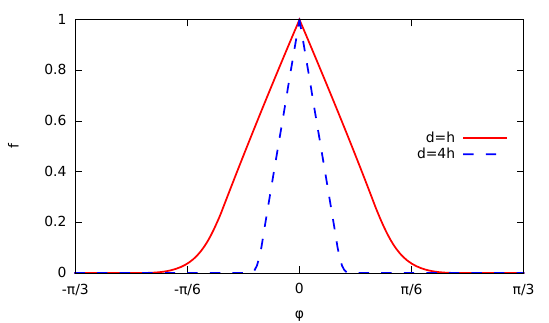}

\caption{\label{fig:10} The function $f(\varphi)$ Eq.~\ref{eq:fofphi} 
for $l=h$ and $d=h$ (red solid line) as well as $d=4h$
(blue solid line).
}

\end{figure}
For the device shown in Fig.~\ref{fig:2Dapprox} 
we can derive $f\left(\varphi\right)$
from the sketch shown in Fig.~\ref{fig:Sketch}. Considering an electron
starting from the FM at an angle $\varphi$ on a vertical line with distance
$x$ from a vertical line in the metallic electrode (M), it can only reach
that line if it comes from the fraction $h'=h-x\tan\varphi$, i.e.
the mean probability to reach the line is 
$\frac{h'}{h}=\left(1-\frac{x}{h}\tan\left|\varphi\right|\right)$
if $\left|\varphi\right|\leq\varphi_{m}=\arctan\frac{h}{x}$ and otherwise
zero. Averaging over the length $l$ of the FM and the metallic electrode
results in 
\begin{eqnarray}\label{eq:fofphi}
f\left(\varphi\right) & = & \int_{d}^{d+2l}dx\frac{l-\left|x-d-l\right|}{l^{2}}\\
 &  & \cdot\left(1-\frac{x}{h}\tan\left|\varphi\right|\right)\Theta\left(\arctan\frac{h}{x}-\left|\varphi\right|\right),\nonumber 
\end{eqnarray}
where $\Theta$ is the Heaviside step function. $\gamma$ can then
be calculated numerically for specific values of $d$, $l$ and $h$. 
In Fig.~\ref{fig:10} this function $f\left(\varphi\right)$
is shown for $l=h$ and $d=h$ as well as $d=4h$, showing that for
larger value of $d$ the angular dependence becomes more strongly
focused near $\varphi=0$.


\begin{thebibliography}{10}
\bibitem{Bernevig}
B.~A.~Bernevig, T.~L.~Hughes, and S.-C.~Zhang, Quantum Spin Hall Effect 
and Topological Phase Transition in HgTe Quantum Wells, 
\href{http://dx.doi.org/10.1126/science.1133734}{Science {\bf 314}, 1757 (2006)}.

\bibitem{Fu}
L. Fu, C.~L. Kane and E.~J. Mele, Topological Insulators in Three Dimensions,
\href{http://dx.doi.org/10.1103/PhysRevLett.98.106803}{Phys. Rev. Lett. {\bf 98}, 106803 (2007)}.

\bibitem{Koenig}
M.~K\"onig, S.~Wiedmann, C.~Br\"une, A.~Roth, H.~Buhmann,
L.W.~Molenkamp, X.-L.~Qi, and S.-C.~Zhang, Quantum Spin Hall Insulator State in HgTe Quantum Wells 
\href{http://dx.doi.org/10.1126/science.1148047}{Science {\bf 318}, 766 (2007)}.

\bibitem{Hsieh1}
D. Hsieh, D. Qian, L. Wray, Y. Xia, Y. Hor, R.J. Cava, and M.Z. Hasan,
A topological Dirac insulator in a quantum spin Hall phase, 
\href{http://dx.doi.org/10.1038/nature06843}{Nature (London) {\bf 452}, 970 (2008)}.

\bibitem{Chen}Y.~L.~Chen, J.~G.~Analytis, J.-H.~Chu, Z.~K.~Liu,
S.-K.~Mo, X.~L.~Qi, H.~J.~Zhang, D.~H.~Lu, X.~Dai, Z.~Fang,
S.~C.~Zhang, I.~R.~Fisher, Z.~Hussain, and Z.-X.~Shen, Experimental
Realization of a Three-Dimensional Topological Insulator, Bi$_{2}$Te$_{3}$,
\href{http://dx.doi.org/10.1126/science.1173034}{Science {\bf 325}, 178 (2009)}.

\bibitem{Xia} Y.~Xia, D.~Qian, D.~Hsieh, L.~Wray, A.~Pal, H.~Lin,
A.~Bansil, D.~Grauer, Y.~S.~Hor, R.~J.~Cava, and M.~Z.~Hasan,
Observation of a large-gap topological-insulator class with a single
Dirac cone on the surface, \href{http://dx.doi.org/10.1038/nphys1274}{Nature Phys. {\bf 5}, 398 (2009)}.

\bibitem{Hsieh2}
D. Hsieh, Y. Xia, D. Qian, L. Wray, F. Meier, J.H. Dil, J. Osterwalder,
L. Patthey, A.V. Fedorov, H. Lin, A. Bansil, D. Grauer, Y.S. Hor,
R.J. Cava, and M.Z. Hasan, Observation of Time-Reversal-Protected
Single-Dirac-Cone Topological-Insulator States in Bi$_2$Te$_3$ and Sb$_2$Te$_3$,
\href{http://dx.doi.org/10.1103/PhysRevLett.103.146401}{Phys. Rev. Lett. {\bf 103}, 146401 (2009)}.

\bibitem{Kuroda} K.~Kuroda, H.~Miyahara, M.~Ye, S.~V.~Eremeev,
Yu.~M.~Koroteev, E.~E.~Krasovskii, E.~V.~Chulkov, S.~Hiramoto,
C.~Moriyoshi, Y.~Kuroiwa, K.~Miyamoto, T.~Okuda, M.~Arita, K.~Shimada,
H.~Namatame, M.~Taniguchi, Y.~Ueda, and A.~Kimura, Experimental
Verification of PbBi$_{2}$Te$_{4}$ as a 3D Topological Insulator,
\href{http://dx.doi.org/10.1103/PhysRevLett.108.206803}{Phys. Rev. Lett. {\bf 108}, 206803 (2012)}.

\bibitem{Ando} Y.~Ando, Topological Insulator Materials, \href{http://dx.doi.org/10.7566/JPSJ.82.102001}{J. Phys. Soc. Japan {\bf 82}, 102001 (2013)}.

\bibitem{Hsieh} D.~Hsieh, Y.~Xia, D.~Qian, L.~Wray, J.~H.~Dil,
F.~Meier, J.~Osterwalder, L.~Patthey, J.~G.~Checkelsky, N.~P.~Ong,
A.~V.~Fedorov, H.~Lin, A.~Bansil, D.~Grauer, Y.~S.~Hor, R.~J.~Cava,
and M.~Z.~Hasan, A tunable topological insulator in the spin helical
Dirac transport regime, \href{http://dx.doi.org/10.1038/nature08234}{Nature (London) {\bf 460}, 1101 (2009)}.

\bibitem{Bruene} C.~Br\"une, A.~Roth, H.~Buhmann, E.~M.~Hankiewicz,
L.~W.~Molenkamp, J.~Maciejko, X.-L.~Qi and S.-C.~Zhang, Spin
polarization of the quantum spin Hall edge states, \href{http://dx.doi.org/10.1038/nphys2322}{Nature Phys. {\bf 8},  485  (2012)}.

\bibitem{Pan:PRL106} Z.-H.~Pan, E.~Vescovo, A.~V.~Fedorov, D.~Gardner,
Y.~S.~Lee, S.~Chu, G.~D.~Gu, and T.~Valla, Electronic Structure
of the Topological Insulator Bi$_{2}$Se$_{3}$ Using Angle-Resolved
Photoemission Spectroscopy: Evidence for a Nearly Full Surface Spin
Polarization, \href{http://link.aps.org/doi/10.1103/PhysRevLett.106.257004}{Phys. Rev. Lett. {\bf 106}, 257004 (2011)}.

\bibitem{Tanaka} Y. Tanaka, T. Yokoyama, and N. Nagaosa, Manipulation
of the Majorana Fermion, Andreev Reflection, and Josephson Current
on Topological Insulators, \href{http://link.aps.org/doi/10.1103/PhysRevLett.103.107002}{Phys. Rev. Lett. {\bf   103}, 107002 (2009)}.

\bibitem{Mondal} S. Mondal, D. Sen, K. Sengupta, and R. Shankar,
Tuning the Conductance of Dirac Fermions on the Surface of a Topological
Insulator, \href{http://link.aps.org/doi/10.1103/PhysRevLett.104.046403}{Phys. Rev. Lett. {\bf 104}, 046403 (2010)}.

\bibitem{Linder} J. Linder, Y. Tanaka, T. Yokoyama, A. Sudbo, and
N. Nagaosa, Unconventional Superconductivity on a Topological Insulator,
\href{http://link.aps.org/doi/10.1103/PhysRevLett.104.067001}{Phys. Rev. Lett. {\bf 104}, 067001 (2010)}.

\bibitem{Garate} I. Garate and M. Franz, Inverse Spin-Galvanic Effect
in the Interface between a Topological Insulator and a Ferromagnet,
\href{http://link.aps.org/doi/10.1103/PhysRevLett.104.146802}{Phys. Rev. Lett. {\bf 104}, 146802 (2010)}.

\bibitem{Yokoyama} T. Yokoyama, Y. Tanaka, and N. Nagaosa, Anomalous
magnetoresistance of a two-dimensional ferromagnet/ferromagnet junction
on the surface of a topological insulator, \href{http://link.aps.org/doi/10.1103/PhysRevB.81.121401}{Phys. Rev. B {\bf 81}, 121401 (2010)}.

\bibitem{BlackSchaffer} A.M. Black-Schaffer and J. Linder, Majorana
fermions in spin-orbit-coupled ferromagnetic Josephson junctions,
\href{http://link.aps.org/doi/10.1103/PhysRevB.84.180509}{Phys. Rev. B {\bf 84}, 180509(R) (2011)}.

\bibitem{Krueckl} V.~Krueckl and K.~Richter, Switching Spin and
Charge between Edge States in Topological Insulator Constrictions,
\href{http://link.aps.org/doi/10.1103/PhysRevLett.107.086803}{Phys. Rev. Lett. {\bf 107}, 086803 (2011)}.

\bibitem{Kong} B. D. Kong, Y. G. Semenov, C. M. Krowne, and K. W.
Kim, Unusual magnetoresistance in a topological insulator with a single
ferromagnetic barrier,
\href{http://dx.doi.org/10.1063/1.3600330}{Appl. Phys. Lett. {\bf 98}, 243112
  (2011)}.

\bibitem{Salehi} M.​ Salehi, M​​. Alidoust, Y​.​Rahnavard, G​.​Rashedi,
In-plane magnetoresistance on the surface of topological insulator
\href{http://dx.doi.org/10.1016/j.physe.2010.11.026}{Physica E {\bf 43}, 966 (2011)}.

\bibitem{Vali} R.~Vali and M.~Vali, Tunneling conductance and magnetoresistance
in topological insulator Fi/I/Fi/d-wave superconductor junctions,
\href{http://dx.doi.org/10.1063/1.4766286}{ J. Appl. Phys. {\bf 112}, 103919 (2012)}.

\bibitem{PGGD} T.~Paananen, H.~Gerber, M.~G\"otte, and T.~Dahm,
Appearance of flat surface bands in three-dimensional topological
insulators in a ferromagnetic exchange field, \href{http://stacks.iop.org/1367-2630/16/033019}{New J. Phys. {\bf 16}, 033019 (2014)}.

\bibitem{Taguchi} K.~Taguchi, T.~Yokoyama, and Y.~Tanaka, 
Giant magnetoresistance in the junction of two ferromagnets on the surface of diffusive topological insulators,
\href{http://link.aps.org/doi/10.1103/PhysRevB.89.085407}{Phys. Rev. B {\bf 89}, 085407 (2014)}.

\bibitem{Li:NPhys10} R.~Li, J.~Wang, X.-L.~Qi, and S.-C.~Zhang,
Dynamical axion field in topological magnetic insulators, \href{http://dx.doi.org/10.1038/nphys1534}{Nature Phys. {\bf 6},  284  (2010)}.

\bibitem{Zhang:NPhys09} H.~Zhang, C.-X.~Liu, X.-L.~Qi, X.~Dai,
Z.~Fang, and S.-C.~Zhang, Topological insulators in Bi$_{2}$Se$_{3}$,
Bi$_{2}$Te$_{3}$ and Sb$_{2}$Te$_{3}$ with a single Dirac cone
on the surface, \href{http://dx.doi.org/10.1038/nphys1270}{Nature Phys. {\bf 5},  438  (2009)}.

\bibitem{Shan} W.-Y.~Shan, H.-Z.~Lu, and S.-Q.~Shen, Effective
continuous model for surface states and thin films of three-dimensional
topological insulators, \href{http://dx.doi.org/10.1088/1367-2630/12/4/043048}{New J. Phys. {\bf 12}, 043048 (2010)}.

\bibitem{Liu:PRB10} C.-X.~Liu, X.-L.~Qi, H.J.~Zhang, X.~Dai,
Z.~Fang, and S.-C.~Zhang, Model Hamiltonian for topological insulators,
\href{http://link.aps.org/doi/10.1103/PhysRevB.82.045122}{Phys. Rev. B {\bf 82}, 045122 (2010)}.

\bibitem{Hao}L.~Hao and T.~K.~Lee, Surface spectral function in
the superconducting state of a topological insulator, \href{http://dx.doi.org/10.1103/PhysRevB.83.134516}{Phys. Rev. B {\bf 83}, 134516 (2011)}.

\bibitem{Fu2} L.~Fu, Hexagonal Warping Effects in the Surface States of the Topological Insulator Bi$_2$Te$_3$
\href{http://link.aps.org/doi/10.1103/PhysRevLett.103.266801}{Phys. Rev. Lett. {\bf 103}, 266801 (2009)}.

\bibitem{Okamoto} H.~Okamoto, The Bi-Se (Bismuth-Selenium) System,
  \href{http://dx.doi.org/10.1007/BF02646366}{J. Phase Equilib. {\bf 15}, 195 (1994)}.

\bibitem{Lind} H.~Lind, S.~Lidin, and U.~H\"aussermann, Structure and bonding
  properties of (Bi$_2$Se$_3$)$_m$(Bi$_2$)$_n$ stacks by first-principles
  density functional theory, 
\href{http://dx.doi.org/10.1103/PhysRevB.72.184101}{Phys. Rev. B {\bf 72}, 184101 (2005)}.

\bibitem{Zhang:APL09}G.~Zhang, H.~Qin, J.~Teng, J.~Guo, Q.~Guo,
X.~Dai, Z.~Fang, and K.~Wu, Quintuple-layer epitaxy of thin films
of topological insulator Bi$_{2}$Se$_{3}$, \href{http://dx.doi.org/10.1063/1.3200237}{Appl. Phys. Lett. {\bf 95}, 053114 (2009)}.

\bibitem{YiZhang} Yi~Zhang \textit{et~al.}, Crossover of the three-dimensional
topological insulator Bi$_{2}$Se$_{3}$ to the two-dimensional limit,
\href{http://dx.doi.org/10.1038/nphys1689}{Nature Phys. {\bf 6}, 584 (2010)}.

\bibitem{PD} T.~Paananen and T.~Dahm, Magnetically robust topological
edge states and flat bands, \href{http://link.aps.org/doi/10.1103/PhysRevB.87.195447}{Phys. Rev. B {\bf       87}, 195447 (2013)}.

\bibitem{Prange} R.E.~Prange, Tunneling from a Many-Particle Point
of View, \href{http://dx.doi.org/10.1103/PhysRev.131.1083}{Phys. Rev. {\bf 131}, 1083 (1963)}.

\bibitem{Mahan} G.~D.~Mahan, \textit{Many-Particle Physics}, (2nd
ed, Plenum Press, New York 1990).

\bibitem{Silvestrov}P.~G.~Silvestrov, P.~W.~Brouwer, and E.~G.~Mishchenko,
Spin and charge structure of the surface states in topological insulators,
\href{http://dx.doi.org/10.1103/PhysRevB.86.075302}{Phys. Rev. B {\bf 86}, 075302 (2012)}.

\bibitem{Pan:PRB88}Z.-H.~Pan, E.~Vescovo, A.~V.~Fedorov, G.~D.~Gu,
and T.~Valla, Persistent coherence and spin polarization of topological
surface states on topological insulators, \href{http://dx.doi.org/10.1103/PhysRevB.88.041101}{Phys. Rev. B {\bf 88}, 041101(R) (2013)}.

\bibitem{Yazyev}O.~V.~Yazyev, J.~E.~Moore, and S. G. Louie, Spin
Polarization and Transport of Surface States in the Topological Insulators
Bi$_{2}$Se$_{3}$ and Bi$_{2}$Te$_{3}$ from First Principles, \href{http://dx.doi.org/10.1103/PhysRevLett.105.266806}{Phys. Rev. Lett. {\bf 105}, 266806 (2010)}.

\bibitem{Wang}X.~Wang, G.~Bian,~T.~Miller, and T.-C.~Chiang,
Topological spin-polarized electron layer above the surface of Ca-terminated
Bi$_{2}$Se$_{3}$, \href{http://dx.doi.org/10.1103/PhysRevB.87.035109}{Phys. Rev. B {\bf 87}, 035109 (2013)}.

\bibitem{Sanchez} J. S\'anchez-Barriga et al, Photoemission of Bi$_{2}$Se$_{3}$
with Circularly Polarized Light: Probe of Spin Polarization or Means
for Spin Manipulation?, \href{http://dx.doi.org/10.1103/PhysRevX.4.011046}{Phys. Rev. X {\bf 4}, 011046 (2014)}.

\bibitem{Jozwiak}C.~Jozwiak, Y.~L.~Chen, A.~V.~Fedorov, J.~G.~Analytis,
C.~R.~Rotundu, A.~K.~Schmid, J.~D.~Denlinger, Y.-D.~Chuang,
D.-H.~Lee, I.~R.~Fisher, R.~J.~Birgeneau, Z.-X.~Shen, Z.~Hussain,
and A.~Lanzara, Widespread spin polarization effects in photoemission
from topological insulators, \href{http://dx.doi.org/10.1103/PhysRevB.84.165113 }{Phys. Rev. B {\bf 84}, 165113 (2011)}.

\bibitem{Souma:PRL109}S.~Souma, M.~Komatsu, M.~Nomura, T.~Sato,
A.~Takayama, T.~Takahashi, K.~Eto, K.~Segawa, and Y.~Ando, Spin
Polarization of Gapped Dirac Surface States Near the Topological Phase
Transition in TlBi(S$_{1-x}$Se$_{x}$)$_{2}$, \href{http://dx.doi.org/10.1103/PhysRevLett.109.186804}{Phys. Rev. Lett. {\bf 109}, 186804 (2012)}.

\bibitem{Souma:PRL106}S.~Souma, K.~Kosaka, T.~Sato, M.~Komatsu,
A.~Takayama, T.~Takahashi, M.~Kriener, K.~Segawa, and Y.~Ando,
Direct Measurement of the Out-of-Plane Spin Texture in the Dirac-Cone
Surface State of a Topological Insulator, \href{http://dx.doi.org/10.1103/PhysRevLett.106.216803}{Phys. Rev. Lett. {\bf 106}, 216803 (2011)}.

\bibitem{Sato}T.~Sato, K.~Segawa, H.~Guo, K.~Sugawara, S.~Souma,
T.~Takahashi, and Y.~Ando, Direct Evidence for the Dirac-Cone Topological
Surface States in the Ternary Chalcogenide TlBiSe$_{2}$, \href{http://dx.doi.org/10.1103/PhysRevLett.105.136802}{Phys. Rev. Lett. {\bf 105}, 136802 (2010)}.

\bibitem{Pauly}C.~Pauly, G.~Bihlmayer, M.~Liebmann, M.~Grob,
A.~Georgi, D.~Subramaniam, M.~R.~Scholz, J.~SÃ¡nchez-Barriga,
A.~Varykhalov, S.~BlÃŒgel, O.~Rader, and M.~Morgenstern, Probing
two topological surface bands of Sb$_{2}$Te$_{3}$ by spin-polarized
photoemission spectroscopy, \href{http://dx.doi.org/10.1103/PhysRevB.86.235106}{Phys. Rev. B {\bf 86}, 235106 (2012)}.

\bibitem{Herdt} A.~Herdt, L.~Plucinski, G.~Bihlmayer, G.~Mussler,
S.~D\"oring, J.~Krumrain, D.~Gr\"utzmacher, S.~Bl\"ugel, and C.~M.~Schneider,
Spin-polarization limit in Bi$_{2}$Te$_{3}$ Dirac cone studied by
angle- and spin-resolved photoemission experiments and ab initio calculations,
\href{http://dx.doi.org/10.1103/PhysRevB.87.035127}{Phys. Rev. B {\bf 87}, 035127 (2013)}.

\bibitem{Scholz} However, a spin polarization of 75\%-80\% in Bi$_{2}$Te$_{3}$
has been reported in M.~R.~Scholz et al, Reversal of the Circular
Dichroism in Angle-Resolved Photoemission from Bi$_{2}$Te$_{3}$,
\href{http://link.aps.org/doi/10.1103/PhysRevLett.110.216801}{Phys. Rev. Lett. {\bf 110}, 216801 (2013)}.

\bibitem{Nomura}M.~Nomura, S.~Souma, A.~Takayama, T.~Sato, T.~Takahashi,
K.~Eto, K.~Segawa, and Y.~Ando, Relationship between Fermi surface
warping and out-of-plane spin polarization in topological insulators:
A view from spin- and angle-resolved photoemission, \href{http://dx.doi.org/10.1103/PhysRevB.89.045134}{Phys. Rev. B {\bf 89}, 045134 (2014)}.

\bibitem{Souma:PRL108}S.~Souma, K.~Eto, M.~Nomura, K.~Nakayama,
T.~Sato, T.~Takahashi, K.~Segawa, and Y.~Ando, Topological Surface
States in Lead-Based Ternary Telluride Pb(Bi$_{1-x}$Sb$_{x}$)$_{2}$Te$_{4}$,
\href{http://dx.doi.org/10.1103/PhysRevLett.108.116801}{Phys. Rev. Lett. {\bf 108}, 116801 (2012)}.

\bibitem{Yamamoto} M.~Yamamoto, T.~Ishikawa, T.~Taira, G.-F.~Li,
K.~Matsuda, T.~Uemura, Effect of defects in Heusler alloy thin films
on spin-dependent tunnelling characteristics of Co$_{2}$MnSi/MgO/Co$_{2}$MnSi
and Co$_{2}$MnGe/MgO/Co$_{2}$MnGe magnetic tunnel junctions , \href{http://dx.doi.org/10.1088/0953-8984/22/16/164212}{J. Phys.:  Condens. Matter {\bf 22} 164212 (2010)}.

\bibitem{Drewello} V.~Drewello, D.~Ebke, M.~Sch\"afers, Z.~Kugler,
G.~Reiss, A.~Thomas, Magnon excitation and temperature dependent
transport properties in magnetic tunnel junctions with Heusler compound
electrodes, \href{http://dx.doi.org/10.1063/1.3669913}{J. Appl. Phys.  {\bf 111}, 07C701 (2012)}.

\bibitem{Koenig1} M.~K\"onig, H.~Buhmann, L.~W.~Molenkamp, T.~Hughes,
C.-X.~Liu, X.-L.~Qi, and S.-C.~Zhang, The Quantum Spin Hall Effect: Theory and Experiment
\href{http://dx.doi.org/10.1143/JPSJ.77.031007}{J. Phys. Soc. Japan {\bf 77}, 031007 (2008)}.

\bibitem{Koenig2} M.~K\"onig, M.~Baenninger, A.~G.~F.~Garcia, N.~Harjee, 
B.~L.~Pruitt, C.~Ames, P.~Leubner, C.~Br\"une, H.~Buhmann, L.~W.~Molenkamp, 
and D.~Goldhaber-Gordon, Spatially Resolved Study of Backscattering in the 
Quantum Spin Hall State,
\href{http://dx.doi.org/10.1103/PhysRevX.3.021003}{Phys. Rev. X {\bf 3}, 021003 (2013)}.

\bibitem{Kubota} H.~Kubota, Y.~Ando, T.~Miyazaki, G.~Reiss, H.~Br\"uckl, 
W.~Schepper, J.~Wecker, G.~Gieres, Size dependence of switching field of 
magnetic tunnel junctions down to 50 nm scale, 
\href{http://dx.doi.org/10.1063/1.1588357}{J. Appl. Phys. {\bf 94}, 2028 (2003)}.

\bibitem{Wojcik} P.~W\'ojcik, J.~Adamowski, B.~J.~Spisak, and M.~Wooszyn,
 \href{http://dx.doi.org/10.1063/1.4868691}{J. Appl. Phys.  {\bf 115}, 104310 (2014)}.

\bibitem{Yoh} K.~Yoh, Z.~Cui, K.~Konishi, M.~Ohno, K.~Blekker, W.~Prost, F.-J.~Tegude,
and J.-C.~Harmand, An InAs Nanowire Spin Transistor with Subthreshold Slope of
20mV/dec, \href{http://dx.doi.org/10.1109/DRC.2012.6256935}{Proceedings of 
IEEE 70th Annual} Device Research Conference, p.~79 (2012).

\bibitem{Parkin} S.~S.~Parkin, C.~Kaiser, A.~Panchula, P.~M.~Rice,
B.~Hughes, M.~Samant, S.-H.~Yang, Giant tunnelling magnetoresistance
at room temperature with MgO (100) tunnel barriers, \href{http://dx.doi.org/10.1038/nmat1256}{Nature   Mater. {\bf 3}, 862 (2004)}.

\bibitem{Yuasa} S.~Yuasa, T.~Nagahama, A.~Fukushima, Y.~Suzuki,
K.~Ando, Giant room-temperature magnetoresistance in single-crystal
Fe/MgO/Fe magnetic tunnel junctions,
\href{http://dx.doi.org/10.1038/nmat1257}{Nature Mater. {\bf 3}, 868 (2004)}.

\bibitem{Ikeda} S.~Ikeda, J.~Hayakawa, Y.~Ashizawa, Y.~M.~Lee, K.~Miura, 
H.~Hasegawa, M.~Tsunoda, F.~Matsukura, and H.~Ohno, Tunnel magnetoresistance
of 604\% at 300K by suppression of Ta diffusion in CoFeB/MgO/CoFeB 
pseudo-spin-valves annealed at high temperature,
\href{http://dx.doi.org/10.1063/1.2976435}{Appl. Phys. Lett. {\bf 93},
 082508 (2008)}.

\bibitem{Liu} H.-X.~Liu, Y.~Honda, T.~Taira, K.~Matsuda, M.~Arita,
T.~Uemura, M.~Yamamoto, Giant tunneling magnetoresistance in epitaxial
Co$_{2}$MnSi/MgO/Co$_{2}$MnSi magnetic tunnel junctions by half-metallicity
of Co$_{2}$MnSi and coherent tunneling,
\href{http://dx.doi.org/10.1063/1.4755773}{Appl. Phys. Lett.  {\bf 101},
  132418 (2012)}.

\end{thebibliography}
\end{document}